\documentclass{article}

\usepackage{arxiv}

\usepackage[utf8]{inputenc} % allow utf-8 input
\usepackage[T1]{fontenc}    % use 8-bit T1 fonts
\usepackage{hyperref}       % hyperlinks
\usepackage{url}            % simple URL typesetting
\usepackage{booktabs}       % professional-quality tables
\usepackage{amsfonts}       % blackboard math symbols
\usepackage{nicefrac}       % compact symbols for 1/2, etc.
\usepackage{microtype}      % microtypography
\usepackage{lipsum}
\usepackage{graphicx}

\graphicspath{ {./images/} }

\usepackage[caption=false,font=normalsize,labelfont=sf,textfont=sf]{subfig}
\usepackage{amsmath}
\usepackage{siunitx}

\title{Synergizing Deep Learning and Full-Waveform Inversion: Bridging Data-Driven and Theory-Guided Approaches for Enhanced Seismic Imaging}

\author{
  Christopher Zerafa\\
  Department of Geosciences\\
  University of Malta\\
  Msida, Malta\\
  \texttt{christopher.zerafa.08@um.edu.mt} \\
   \And
   Pauline Galea \\
  Department of Geosciences\\
  University of Malta\\
  Msida, Malta\\
  \texttt{pauline.galea@um.edu.mt} \\
   \AND
   Cristiana Sebu \\
  Department of Geosciences\\
  University of Malta\\
  Msida, Malta\\
   \texttt{cristiana.sebu@um.edu.mt} 
}

\begin{document}
\maketitle
\begin{abstract}
    
Seismic imaging and subsurface characterization play vital roles in understanding Earth's geological properties. This review paper explores the integration of deep learning techniques with full-waveform inversion (FWI), a critical method for estimating subsurface properties using seismic data. The paper is structured into five chapters that cover the theoretical foundations, applications, challenges, limitations, and future research directions of this integration.
	
Chapter 1 provides an introduction to the motivation and objectives behind the integration of deep learning and FWI. Chapter 2 establishes the fundamentals of FWI, elucidating its mathematical principles, forward and inverse problems, and workflow components. Chapter 3 offers a comprehensive introduction to deep learning, including neural networks, activation functions, optimization algorithms, and regularization techniques.
	
Chapter 4 delves into specific applications of deep learning in geophysics, highlighting its utilization in various domains like seismology, geology, and hydrogeology. Notable applications include velocity estimation, seismic deconvolution, and tomography, showcasing how deep learning enhances subsurface characterization.
	
Chapter 5 addresses critical challenges and potential limitations in this integration, such as model complexity, data quality, and generalization. The chapter outlines future research directions, including hybrid models, generative models, uncertainty quantification, and physics-informed learning.
	
In conclusion, the synergy of deep learning and FWI holds immense potential to revolutionize seismic imaging and subsurface characterization. This integration, while posing challenges, offers pathways to more accurate, efficient, and reliable subsurface property estimation. By addressing challenges and pursuing innovative research, this integration promises to reshape geophysics and contribute to a comprehensive understanding of Earth's subsurface properties.

\end{abstract}

\section{Introduction}

Full Waveform Inversion (FWI) has emerged as a powerful technique in the field of seismic imaging and exploration, revolutionizing our ability to accurately reconstruct subsurface structures and properties. By iteratively minimizing the misfit between observed and modeled seismic data, FWI promises high-resolution reconstructions that have implications spanning from oil and gas reservoir characterization to earthquake hazard assessment. However, despite its immense potential, FWI encounters numerous challenges that hinder its widespread applicability, including sensitivity to initial models, computational intensity, and the requirement of dense data coverage.

In recent years, the advent of Deep Learning (DL) methodologies has ushered in a paradigm shift in various scientific and engineering domains. DL techniques, such as Convolutional Neural Networks (CNNs) and Recurrent Neural Networks (RNNs), have demonstrated exceptional capabilities in image recognition, natural language processing, and medical image analysis. When applied to the field of FWI, these methods offer the potential to overcome some of the traditional limitations by introducing innovative data-driven and theory-guided inversion strategies.

This review paper seeks to explore the evolving landscape of FWI through the lens of deep learning techniques, encapsulating a spectrum ranging from data-driven to theory-guided inversion methodologies. We aim to provide a systematic overview of the various approaches that leverage the synergy between FWI and deep learning, highlighting their contributions, challenges, and potential future directions. By amalgamating traditional FWI techniques with the power of deep learning, researchers and practitioners can harness a synergy that might facilitate enhanced inversion accuracy, reduced computational costs, and improved robustness to noise and incomplete data.

The paper is structured as follows: In Section \ref{sec:lit_rev_fwi}, we provide a brief overview of the traditional FWI formulation, highlighting its mathematical underpinnings and challenges. Section \ref{sec:DNN} delves into the fundamental concepts of deep learning, elucidating its different architectures and their applicability to FWI. Subsequently, Section \ref{sec:app_geophysics} presents a comprehensive survey of deep-learning applications to geophysics. In Section \ref{sec:dl_geo_leg_inv} we initial discuss legacy inversion techniques that employed deep learning. In Secion \ref{sec:dl_geo_data}, we then discuss data-driven approaches, where neural networks are employed to directly learn complex relationships between seismic data and subsurface parameters, and finally. In Section \ref{sec:dl_geo_theory}, we transition towards theory-guided approaches, exploring methodologies that integrate domain-specific geological constraints and physical laws within deep learning frameworks. Finally, in Section \ref{sec:challanges}, we outline critical challenges, potential limitations, and provide insights into possible future research directions within this evolving field.

In conclusion, the synthesis of Full Waveform Inversion and Deep Learning heralds a new era of seismic imaging, offering the potential to address long-standing challenges and transform the way we extract subsurface information. This review intends to serve as a comprehensive guide for researchers, practitioners, and students in understanding the evolution, nuances, and potential of the synergy between these two dynamic fields. Through a holistic analysis of data-driven and theory-guided approaches, we endeavour to catalyze further innovation, foster cross-disciplinary collaborations, and pave the way for more accurate and efficient subsurface characterizations.

\section{Full Waveform Inversion}\label{sec:lit_rev_fwi}
Full Waveform Inversion (FWI) tries to derive the best velocity model and other lithologic properties (as density, anelastic absorption and anisotropy) of the Earth's subsurface to be consistent with recorded data. An exhaustive search for this ideal model is almost impossible and methods for finding an optimal one describing the data space are necessary. There are two main categories for dealing with this problem: (i) global optimization methods, and (ii) direct solving through linearisation.

% Global optimization methods use stochastic processes to try and find the global minimum of the misfit function \cite{Torn1989, Sen1995}. Three most well-known cases of global methods are Monte Carlo methods \cite{Press1968, Biswas2017}, genetic algorithm \cite{Gerstoft1994, Parker1999, Tran2012} and simulated annealing \cite{Rothman1985, Pullammanappallil1994, Tran2011}. Global optimization methods are all very dependent on a fast forward modelling algorithm as they require large amounts of forward modelling calculations. As computers are getting faster and better, the necessity of keeping the parametrization simple might decline. However, current solutions to the seismic inverse problem have to resort to local optimization. The next section reviews direct solving through linearisation for FWI.

\subsection{FWI as a global optimization method}\label{sec:fwi_global_opt}
Global optimization methods use stochastic processes to try and find the global minimum of the misfit function \cite{Torn1989, Sen1995}. Three most well-known cases of global methods are:
\begin{enumerate}
	\item \textbf{Monte Carlo methods}: These are pure random search methods in which models are drawn stochastically from the total model space, forward modelled and the model with lowest cost function is utilized \cite{Press1968, Biswas2017}.
	\item \textbf{Genetic Algorithm methods}: These are based on analogues of biological evolution where a relatively small selection of models is chosen from the model space \cite{Holland1975}. The best models of these parent selection form new child models by cross-over and mutation of the parameters describing the model. The children will then replace the weakest models in the selection. By such iterative steps, the selection of models will gather towards an optimal model \cite{Gerstoft1994, Parker1999, Tran2012}.
	\item \textbf{Simulated Annealing methods}: These are based on analogues of physical annealing processes modelled in statistical mechanics \cite{Kirkpatrick1983, Geman1984}. Model parameters are randomly perturbed and updated model evaluated and assessed whether to be accepted or not. Only better models are propagated forward and the possibility of a perturbation being accepted decreases. If convergence exists, the model will iteratively become better \cite{Rothman1985, Pullammanappallil1994, Tran2011}.
\end{enumerate}

Global optimization methods are all very dependent on a fast forward modelling algorithm as they require large amounts of forward modelling calculations. As computers are getting faster and better, the necessity of keeping the parametrization simple might decline. However, current solutions to the seismic inverse problem have to resort to local optimization.

\subsection[FWI as Local Optimization]{Formulation of FWI as Local Optimization}\label{sec:fwi_local_opt}
The concept of local optimization for FWI was introduced in the 1980’s. \cite{Lailly1983} and \cite{Tarantola1984a} cast the exploding-reflector concept of \cite{Claerbout1971, Claerbout1976} as a local optimization problem which aims to minimise in a least-squares sense the misfit between recorded and modelled data \cite{Virieux2009}. The problem is set in the time-domain as follows: set a forward propagation field to model the observed data, back propagate the misfit between the modelled and the observed data, cross-correlate both fields at each point in space to derive a correction, and do least squares minimization of the residuals iteratively. This outline forms the basis of this technique to this day.

\cite{Gauthier1986} numerically demonstrate a local optimization FWI approach using two-dimensional synthetic data examples. A single diffracting point on a homogeneous model was used to illustrate the importance of proper sampling of the subsurface. Furthermore, this model was used to show that the free surface adds an extra complexity to the problem and increases the non-linearity of the inversion. FWI with or without free-surface multiple modelling is an active area of research to this day \cite{Komatitsch2002,Bergen2019}. 
% A homogeneous model with two point diffractors of equal amplitudes was used to demonstrate the importance of preconditioning the gradient to accelerate the convergence and correctly recover amplitudes. Another synthetic model with a circular perturbation of a diameter larger than the wavelength showed the importance of the accuracy in the starting model. When the starting model falls within a global minimum, the predicted data converges to a true solution, otherwise the inversion results were cycle-skipped and convergence occurs within local minima.

\subsection{Applications in Time and Frequency}
\begin{figure}[!ht]
	\centering
	\subfloat[Section from real Marmousi velocity model.]{\includegraphics[width=0.45\textwidth]{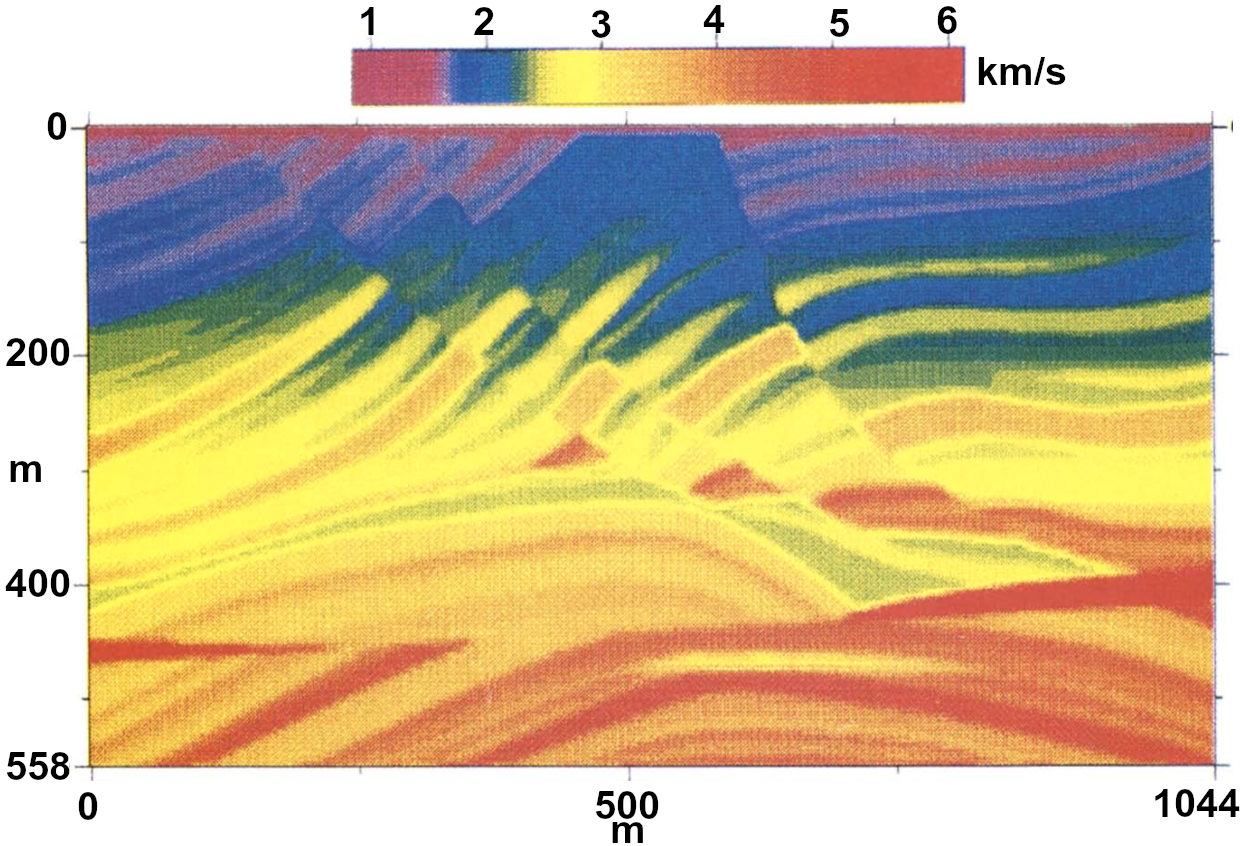}}
	\subfloat[Best estimate using multiscale FWI.]{\includegraphics[width=0.45\textwidth]{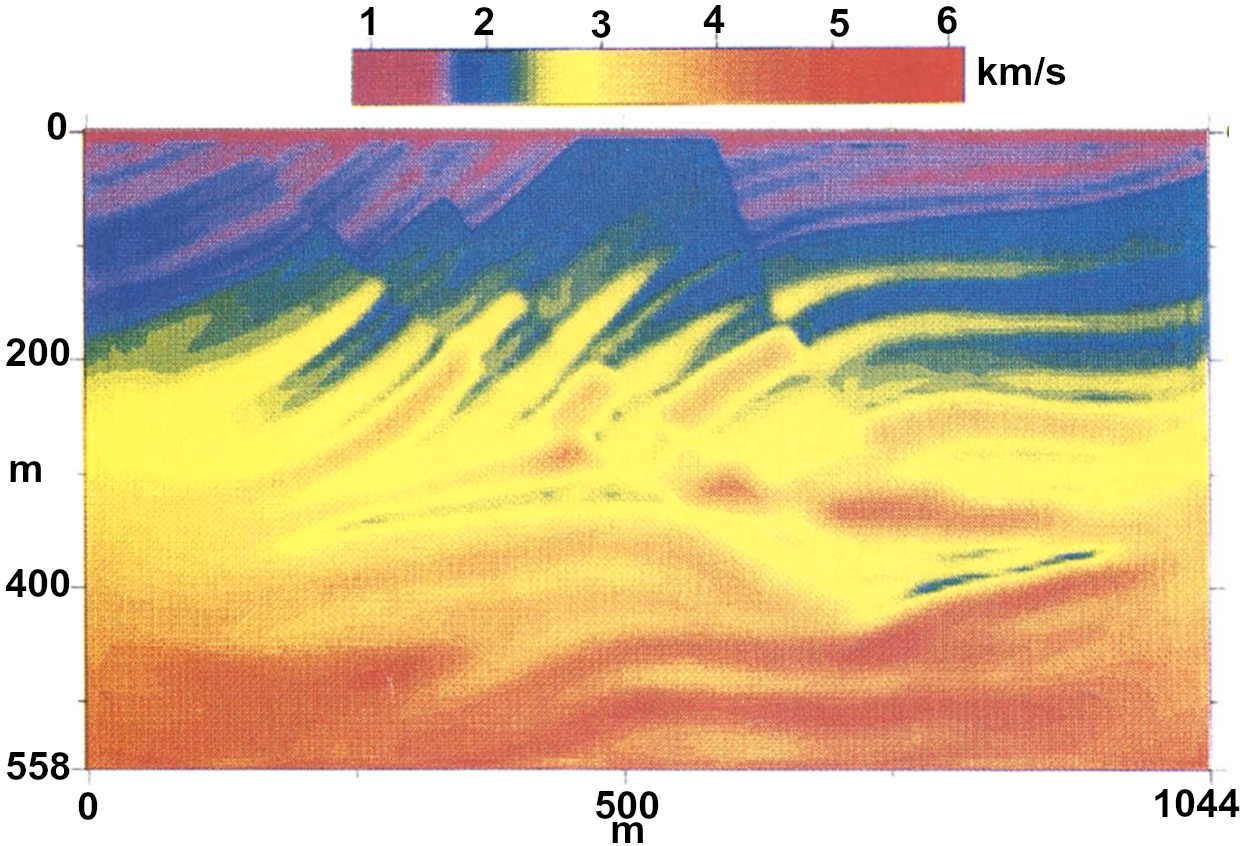}}
	\caption[First practical application of FWI using the Marmousi model]{First practical application of FWI using the Marmousi model. This shows significant improvements for the FWI results as presented by \cite{Bunks1995}.}        
	\label{fig:first_practical_fwi}
\end{figure}

One of the pioneering applications of FWI was presented by \cite{Bunks1995} and is represented in Figure~\ref{fig:first_practical_fwi}. They showed better imaging using a hierarchical multi-scale approach on the Marmousi synthetic model. This strategy initially inverts for low-frequency components where there are fewer local minima, and those that exist are sparser than if for higher frequencies. However, decomposing by scale did not resolve issues of source estimation, source bandwidth and noise \cite{Bunks1995}. In the 1990s, Pratt and his associates proposed FWI via the pseudo-spectral domain \cite{Pratt1990a, Pratt1990b, Pratt1991}. The initial application was to cross-hole data utilizing a finite difference approach and an elastic wave propagator to facilitate the modelling of multi-source data. This was extended to wide-aperture seismic data by \cite{Pratt1996}. Analytically, the time- and frequency-domain problems are equivalent \cite{Virieux2009}. Initial attempts of pseudo-spectral FWI include application to the Marmousi model \cite{Sirgue2004} and land seismic dataset \cite{Operto2004}.

\subsection{Beyond Academic Experiments}\label{sec:lit_rev_beyond_academic_exp}
\begin{figure}
	\centering
	\subfloat[Velocitiy model. Top: Conventional, Bottom: FWI.]{\includegraphics[width=0.45\textwidth]{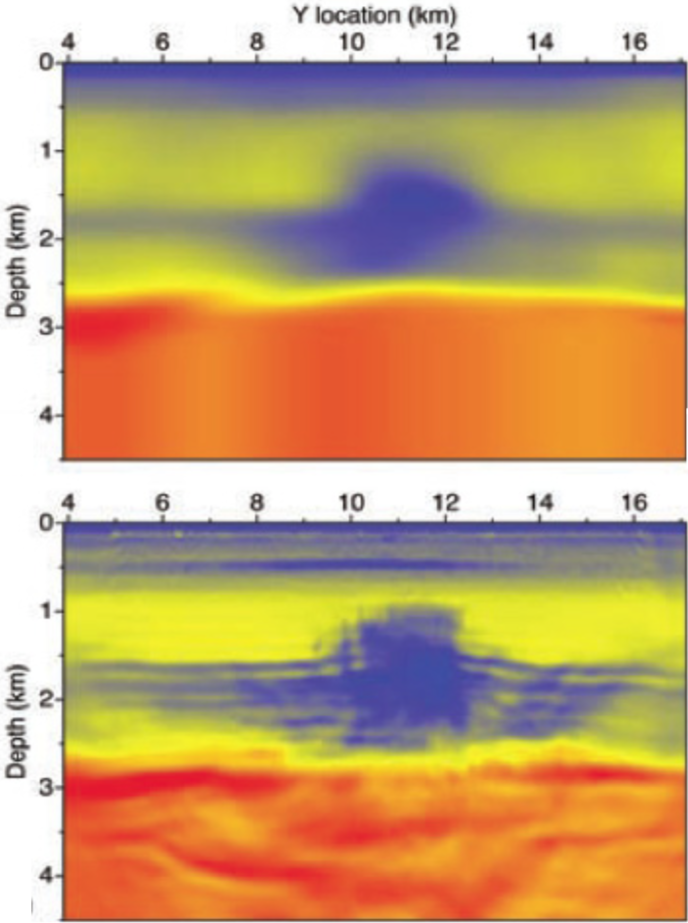}}
	\subfloat[Pre-stack depth migrated section. Top: Conventional, Bottom: FWI.]{\includegraphics[width=0.45\textwidth]{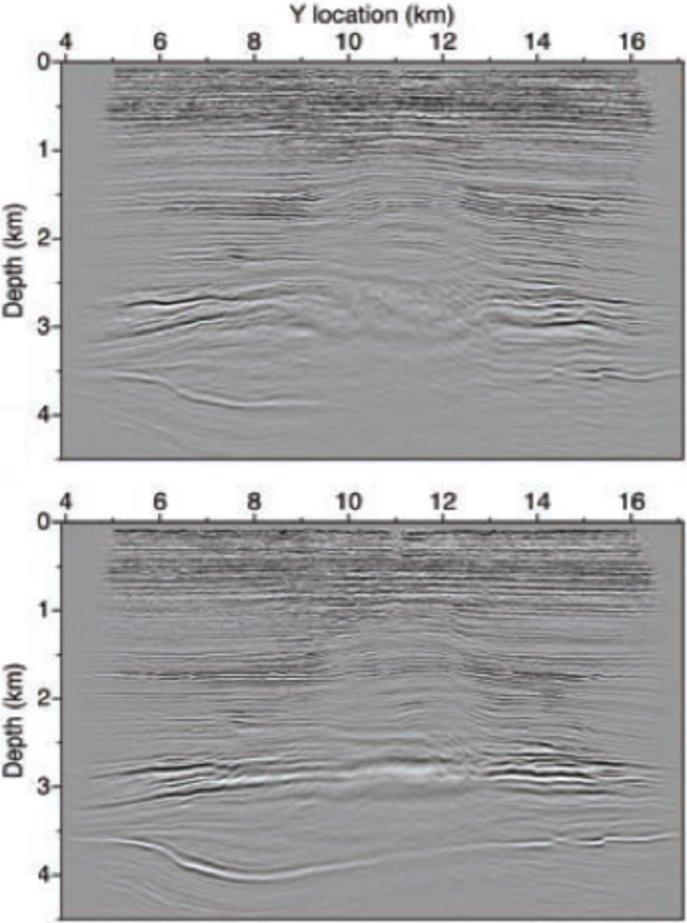}}%
	\caption[Improvements in velocity model and pre-stack depth migrated images obtained through FWI over the Valhall field.]{Improvements in velocity model and pre-stack depth migrated images obtained through FWI over the Valhall field. The FWI updated velocity model demonstrated a network of shallow high-velocity channels and a gas-filled fracture extension from a gas cloud which was not previously identifiable in conventional tomography. The impact is evident in the migrated sections, which show more continues events in otherwise poorly illuminated area. Adapted from \cite{Sirgue2009} and \cite{Sirgue2010}.}        
	\label{fig:fwi_prestack_improvements}
\end{figure}

Theoretically, two-dimensional inversion is only able to explain out-of-plane events by mapping them into in-plane artefacts \cite{Morgan2009}. This meant that FWI restricted to purely academic pursuits \cite{Sirgue2009} and full potential could only be realized if extended to three-dimensions. The first 3D frequency-domain algorithms where developed by \cite{Warner2007} on synthetic datasets, however these used low initial frequencies that are not normally present in real data \cite{Morgan2013}. Examples of this application are demonstrated by \cite{Sirgue2007}, \cite{Ali2007} and \cite{Operto2007}. \cite{Warner2008c} presented the first 3D real data application to a shallow North Sea survey. This improved the resolution of shallow high-velocity channels that resulted in uplifts upon migration. 
% Near-offset, steep reverse faults that segmented the target horizon were clearly identifiable on migrated sections which otherwise were not visible from conventional velocity models.
In Figure~\ref{fig:fwi_prestack_improvements}, \cite{Sirgue2009} demonstrated successful FWI results for a 3D dataset of the Valhall field, Norway. They inverted wide-azimuth ocean-bottom cable data using a sequence of low frequency bands to generate high-resolution velocity models. The updated velocity model demonstrated a network of shallow high-velocity channels and a gas-filled fracture extension from a gas cloud which was not previously identifiable \cite{Sirgue2010}.

\begin{figure}[!ht]
	\centering
	\subfloat[PSDM stack and CDP gather with original orthorhombic model from tomography.]{\includegraphics[width=0.9\textwidth]{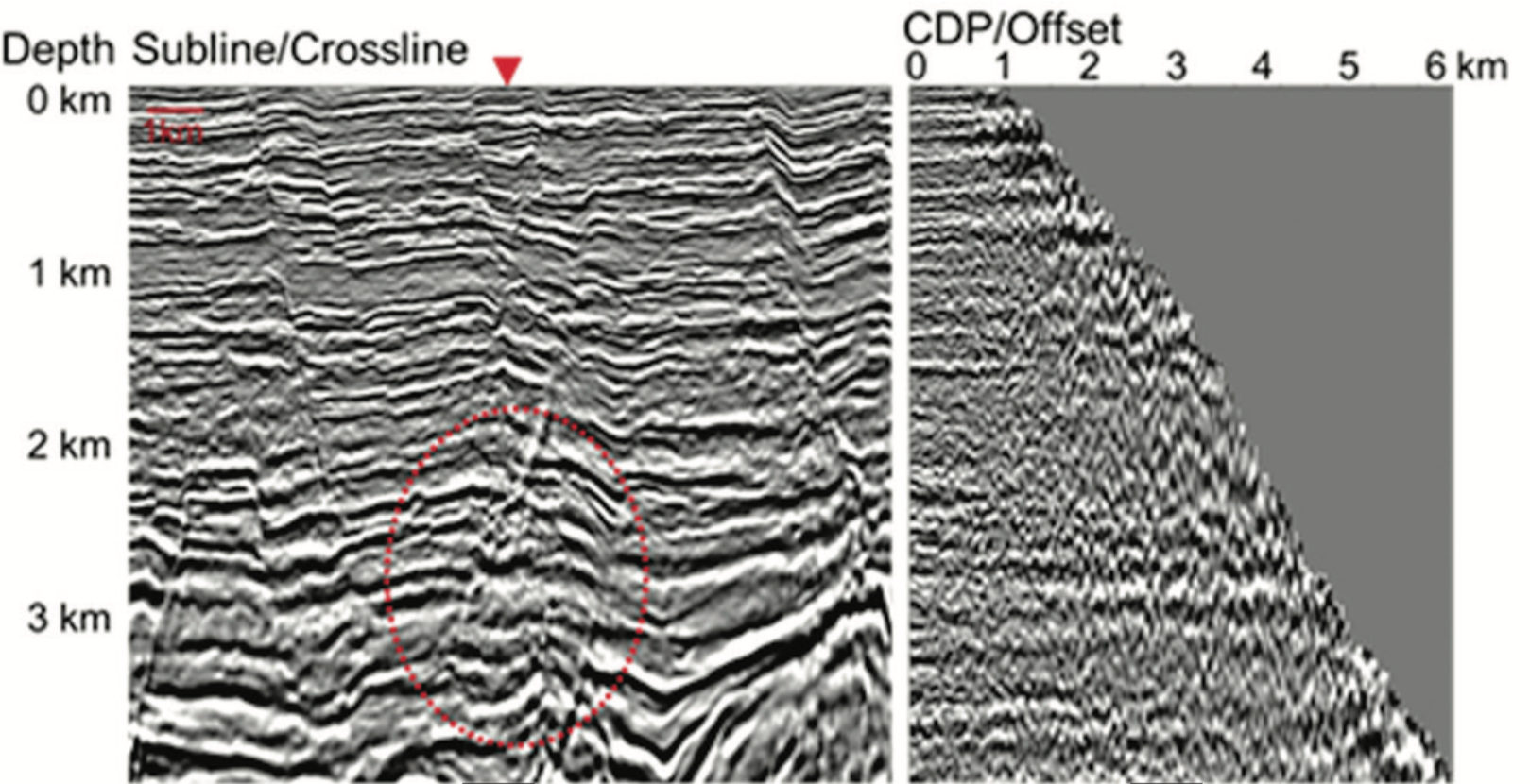}}
    \\
	\subfloat[PSDM stack and CDP gather with orthorhombi FWI update.]{\includegraphics[width=0.9\textwidth]{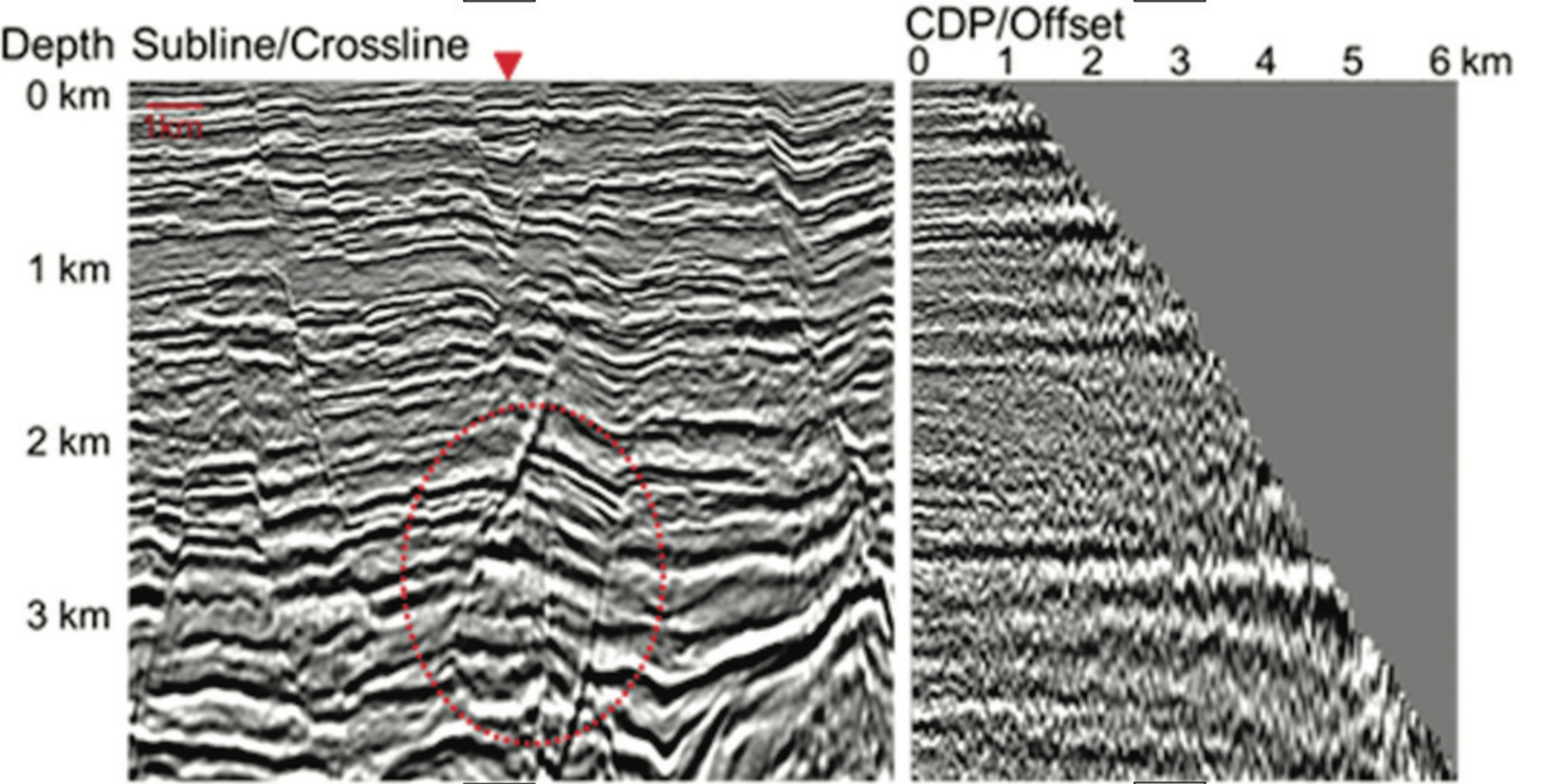}}
	\caption[Imaging improvements obtained through orthorhombic imaging.]{Imaging improvements obtained through orthorhombic imaging. This produces sharp truncations and clearer faults as highlighted by the red dashed ovals, as well as better focussed gathers. Adapted from \cite{Xie2017}.}        
	\label{fig:ortho_fwi}
\end{figure}

\cite{Plessix2010} show results from the application of full waveform inversion to ocean bottom data recorded in the Gulf of Mexico with near-ideal long-offset and wide-azimuth. Their approach was anisotropic and assumed vertical transversely isotropic media with fixed Thomsen’s parameters. The model had better imaging of dips and produced flatter common image gathers in the deep part of the model \cite{Plessix2010}. \cite{Wang2016} developed 3D waveform inversion for orthorhombic media in the acoustic approximation using pseudo-spectral methods. This was found to be stable and produced kinematic accurate pure-mode primary wavefields with an acceptable computational cost. \cite{Xie2017} applied orthorhombic full-waveform inversion for imaging wide-azimuth ocean-bottom-cable data. The results had better azimuthal and polar direction-dependent wave imaging which significantly improved fault imaging -  See Figure~\ref{fig:ortho_fwi}.

A re-occurring theme within this section is the creation of a better approximation to the wavefield propagation within the subsurface; 1D to 2D to 3D discretization, acoustic to anisotropic to elastic to orthorhombic wavefield modelling, with each additional dimension of information resulting in more numerical and computer intensive algorithms \cite{Kumar2012}. Even though computing power has increased dramatically, making FWI more productive, the underlying algorithms are only improving incrementally. Indeed, the next generation of experiments will require changes to acquisition geometry to allow for full-bandwidth and multi-azimuth reconstruction of the wavefield \cite{Morgan2016}.

\section{Deep Neural Networks}\label{sec:DNN}

\subsection{Neural Networks for Inverse Problems}\label{sec:NN_IP}
The mathematical formulation of FWI falls under the more general class of variational inverse problems \cite{Tanaka2003}. The aim is to find a function which is the minimal or the maximal value of a specified functional \cite{Dadvand2006}. Indeed, inverse problems attempt to reconstruct an image $x \in X \subseteq \mathbb{R}^d$ from a set of measurements $y \in Y \subseteq \mathbb{R}^m$ of the form
\begin{equation}\label{eq:inverse_theory}
	y=\Gamma(x)+\epsilon
\end{equation}
where $\Gamma: X \mapsto Y, \Gamma \in \mathbb{R}^{m\times d}$ is the discrete operator and $\epsilon \in Y \subseteq \mathbb{R}^{m}$ is the noise. NN within Machine Learning can be considered to be a set of algorithms of non-linear functional approximations under weak assumptions \cite{Oktem2018}. Namely, when applied to inverse problems, Equation~\ref{eq:inverse_theory} can be re-phrased as the problem of reconstructing a non-linear mapping $\Gamma_\theta^{\dagger}:Y \mapsto X$ satisfying the pseudo-inverse property
\begin{equation}\label{eq:learned_inverse}
	\Gamma_\theta^{-1}(y) \approx x
\end{equation}
where observations $y\in Y$ are related to $x\in X$ as in Equation~\ref{eq:inverse_theory}, and $\theta$ represents the parametrization of pseudo-inverse by the NN learning \cite{Adler2017a}. The loss function defined in Equation~\ref{eq:learned_inverse} is dependent on the type of training data, which is dependent on the learning approach \cite{Adler2017b}.  There are two main classes of learning in Machine Learning: (i) Supervised, and (ii) Unsupervised.

% \subsection{Supervised learning}
In supervised learning, training data are independent distributed random pairs with input $x\in X$ and labelled output $y \in Y$ \cite{Vito2005}. Estimating $\theta$ for Equation~\ref{eq:learned_inverse} can be formulated as minimizing a loss function $\mathcal{J}(\theta)$ which has the following structure \cite{Adler2017a}:
\begin{equation}\label{eq:supervised_learning}
	\mathcal{J}(\theta) := \mathcal{D}(\Gamma_\theta^{-1}(\boldsymbol{y}), \boldsymbol{x})
\end{equation}
where $\mathcal{D}$ is a distance function quantifying the quality of the reconstruction and $\Gamma_\theta^{-1}:Y\mapsto X$ is the pseudo-inverse to be learned \cite{Adler2017a}. A common metric for the distance function is the sum of squared distances, resulting in:
\begin{equation}
\mathcal{J}(\theta) := \left|\left|\Gamma_\theta^{-1}(y)- x\right|\right|_X^2
\end{equation}
Approaching the inverse problem directly via this approach amounts to learn $\Gamma_\theta^{-1}:Y\mapsto X$ from data such that it approximates an inverse of $\Gamma$. In particular, this has successful applications in medical imaging \cite{Xu2012, Lucas2018}, signal processing \cite{Rusu2017, Dokmanic2016} and regularization theory \cite{Meinhardt2017, Romano2016, DelosReyes2017}.

% \subsection{Unsupervised learning}
In unsupervised learning, there exist no input-output labelled pairs and the training data is solely elements of $y\in Y$. The NN is required to learn both the forward problem and inverse problem \cite{Andrychowicz2016}. The loss functional for unsupervised learning is given as:
\begin{equation}
	\mathcal{J}(\theta) := \mathcal{L}\left(\Gamma\left(\Gamma_\theta^{-1}(y)\right),x \right)+\mathcal{S}\left(\Gamma_\theta^{-1}(g)\right)
\end{equation}
where $\mathcal{L}:Y\times X \mapsto\mathbb{R}$ is a suitable affine transformation of the data and $\mathcal{S}:X\mapsto \mathbb{R}$ is the regularization function. Main applications of this learning are to inherent structure and have been proven successful in exploratory data analysis applications such as clustering \cite{Sever2015,Gerdova2002} and dimension reduction \cite{Dolenko2015}.

% \begin{figure}[!ht]
% 	\centering
% 	\subbottom[Supervised learning overview]{\includegraphics[width=0.72\textwidth]{Learning_Type_Supervised.png}}
% 	\subbottom[Unupervised learning overview]{\includegraphics[width=0.72\textwidth]{Learning_Type_Unsupervised.png}}%
% 	\caption[Supervised and unsupervised learning types.]{Supervised and unsupervised learning types. Adapted from \cite{Ma2018}.}        
% 	\label{fig:learning_types}
% \end{figure}

\subsection{Evolution of Neural Networks}\label{sec:evo_NN}
The remaining literature review is restricted to supervised learning approaches using NN as these are more suited for velocity inversion. For a complete review, \cite{Lippmann1987} and \cite{Chentouf1997} provide further detail. 
% This section proceeds with the development of feed-forward networks, overview of \ac{DNN} landscape and architecture type (§\ref{sec:dnn_arch}) and finally investigate applications to Geophysical applications (§\ref{sec:app_geophysics}).

\subsubsection{Early Neural Nets and the Perceptron}\label{sec:early_nn}
The basic ideas of NN date back to the 1940’s and were initially devised by \cite{McCulloch1943} when trying to understand how to map the inner workings of a biological brain into a machine. From a biological aspect, neurons in the brain are interconnected via nerve cells that are involved in the processing and transmitting of chemical and electrical signals \cite{McCulloch1943}. 
% A simplified schematic of the biological brain is illustrated in Figure~\ref{fig:brain_neuron}.

% \begin{figure}[ht!]
% 	\centering
% 	\includegraphics[width=0.7\linewidth]{brain_neuron_2}
% 	\caption[This is the short caption for List of Figures]{Schematic of the biological brain neuron. The main components are the Axons (transmission lines), Dendrites (receptors) and the neuron cell body. These are linked together from previous neurons to form a neural network. From \cite{Preetham2016}.}
% 	\label{fig:brain_neuron}
% \end{figure}

\begin{figure}[ht!]
	\centering
	\includegraphics[width=0.75\linewidth]{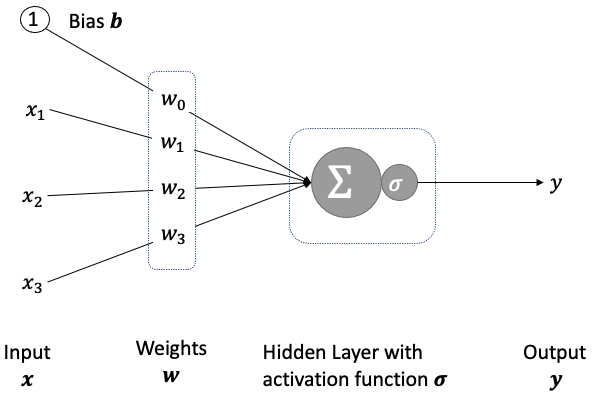}
	\caption[The Single Neuron Perceptron]{The Single Neuron Perceptron. The input values are multiplied by the weights. If the weighted sum of the product satisfies the activation function, the perceptron is activated and ``fires'' a signal. Adapted from \cite{Rosenblatt1958}.}
	\label{fig:perceptron}
\end{figure}

\begin{figure}[ht!]
	\begin{minipage}[c]{0.4\textwidth}
		\centering
		\includegraphics[width=\textwidth]{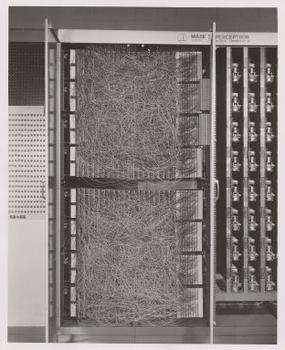}
	\end{minipage}\hfill
	\begin{minipage}[c]{0.55\textwidth}
		\caption[Mark I Perceptron Machine]{The Mark I Perceptron Machine was the first machine used to implement the Perceptron algorithm. The machine was connected to a camera that used 20×20 array of cadmium sulphide photocells to produce a 400-pixel image. To the right is a patch-board that allowed experimentation with different combinations of input features. This was usually wired up randomly to demonstrate the ability of the perceptron to learn. Adapted from \cite{Hecht-Nielsen1990, Bishop2006}.}
		\label{fig:perceptron_mk1}
	\end{minipage}
\end{figure}

Early NN with rudimentary architectures did not learn \cite{McCulloch1943} and the notion of self-organized learning only came about in 1949 by \cite{Hebb1949}. \cite{Rosenblatt1958} extended this idea of learning and proposed the first and simplest neural network – the McCullock-Pitts-Perceptron. As shown in Figure~\ref{fig:perceptron}, this consists of a single neuron with weights and an activation function. The weights are the learned component and determine the contribution of either input $x$ to the output $y$. The activation function $\sigma$ adds a non-linear transform, allowing the neuron to decide if the input is relevant for the paired output. Without an activation function, the neuron would be equivalent to a linear regressor \cite{Minsky2017}. \cite{Rosenblatt1958} used this fundamental architecture to reproduce a functional mapping that classifies patterns that are linearly separable. This machine was an analogue computer that was connected to a camera that used 20×20 array of cadmium sulphide photocells to produce a 400-pixel image. Shown in Figure~\ref{fig:perceptron_mk1}, the McCullock-Pitts-Perceptron had a patch-board that allowed experimentation with different combinations of input features wired up randomly to demonstrate the ability of the perceptron to learn \cite{Hecht-Nielsen1990, Bishop2006}.

Rosenblatt’s perceptron was the first application of supervised learning \cite{Russell2008}. However, \cite{Minsky1969} highlight limitations to the applications of a single perceptron. They also point out that Rosenblatt’s claims that the ``perceptron may eventually be able to learn, make decisions, and translate languages'' were exaggerated. There was no other follow ups on this work by Minsky and Papert, and research on perceptron-style learning machines practically halted \cite{Minsky2017}. 

\subsubsection{Back-Propagation and Hidden Layers}
Efficient error back-propagation in NN networks were described in Linnainmaa’s master thesis \cite{Linnainmaa1970}. This minimizes the errors through gradient descent in the parameter space \cite{Hadamard1907} and allows for explicit minimization of the cost function. Back-propagation permits NN to learn complicated multidimensional functional mappings \cite{Dreyfus1973}. 

The back-propagation formulation lends itself from major developments in dynamic programming throughout the 1960s and 1970s \cite{Kelley1960, Bryson1961, Linnainmaa1976}. A simplified derivation using the chain rule was derived by \cite{Dreyfus1973} and the first NN-specific application was described by \cite{Werbos81}. It was until the mid-1980s that \cite{Rumelhart1986} made back-propagation mainstream for NN through the numerical demonstration of internal representations of the hidden layer. Hidden layers reside in-between input and output layers of the NN.

Back-propagation was no panacea and additional hidden layers did not offer empirical improvements \cite{Schmidhuber2015}. \cite{Kolmogoro1956}, \cite{Hecht-Nielsen1989} and \cite{Hornik1989} pursued development of back-propagation encouraged by the Universal Approximation Theorem. Namely, this theorem states that if enough hidden units are used in a NN layer, this can approximate any multivariate continuous function with arbitrary accuracy \cite{Hecht-Nielsen1989}.
% Numerous improvements to back-propagation have been proposed; least-squares/Gauss-Newton/Levenberg-Marquardt methods \cite{Gauss1809,Newton1687, Levenberg1944}, quasi-newton/Broyden-Fletcher-Goldfarb-Shanno (BFGS) methods \cite{Broyden1965, Fletcher1963}, Partial BFGS \cite{Battiti1992}, conjugate gradient \cite{Hestenes1952}. A full historic recollection of improvements to NN have been described by  \cite{Montavon2012}. 
Although back-propagation theoretically allows for deep problems, it was shown not to work on practical problems \cite{Schmidhuber2015}.

\subsubsection{The Vanishing Gradient and Renaissance of Machine Learning}
The major milestone in NN came about in 1991. Hochreiter’s thesis identified that deep NN suffer from the vanishing or exploding gradient problem \cite{Hochreiter1991}. Gradients computed by back-propagation become very small or very large with added layers, causing convergence to halt or introduce unstable update steps. Solutions proposed to address this challenge included batch normalization \cite{Ioffe2015}, Hessian-free optimisations \cite{Moller1993, Schraudolph2002, Martens2010}, random weight assignment
\cite{Hochreiter1996}, universal sequential search \cite{Levin1973} and weight pruning \cite{LeCun1990}. 

\begin{figure}[ht!]
	\centering
	\includegraphics[width=0.9\linewidth]{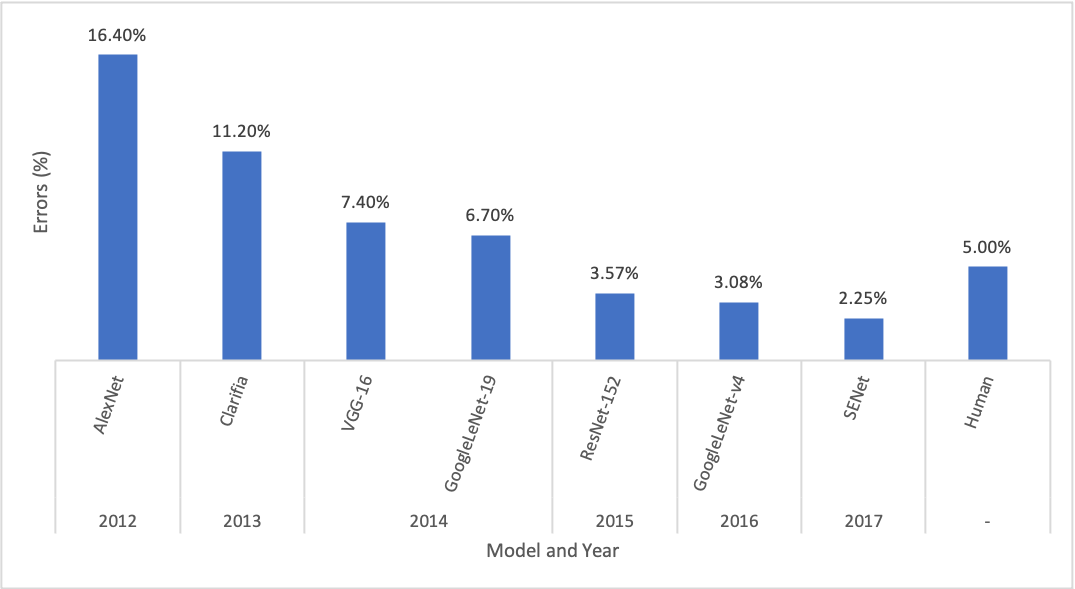}
	\caption[Evolution of accuracy for the ImageNet challenge.]{Evolution of the accuracy for the ImageNet challenge \cite{Krizhevsky2012}. AlexNet won ImageNet in 2012 with 16.4\% error in accuracy. With each year of the competition, the accuracy has been increasing. Sources for these accuracies are Clarifia \cite{Zeiler2014}, VGG-16 \cite{Simonyan2014}, GoogleLeNet-19 \cite{Szegedy2014}, ResNet-152 \cite{He2016}, GoogleLeNet-v4 \cite{Szegedy2016} and SENet \cite{Hu2017}.}
	\label{fig:evolution_imagenet}
\end{figure}

Prior to 2012, NN were apparently an academic pursuit. This changed when AlexNet \cite{Krizhevsky2012} won the ImageNet \cite{Russakovsky2015} visual object recognition by a considerable margin. AlexNet used a deep architecture consisting of eight layers \cite{Krizhevsky2015} and was the only entry employing NN in 2012. All submissions in subsequent years were NN-based \cite{Singh2015} and in 2015, NN surpassed human performance in visual object recognition for the first time \cite{Russakovsky2015} - see Figure~\ref{fig:evolution_imagenet}. AlexNet is undoubtedly a pivotal event that ignited the renaissance in interest around deep learning.

\subsection{Deep Neural Network Architecture Landscape}\label{sec:dnn_arch}
According to \cite{Patterson2017}, three of the most common major architectures are (i) Neural Network, (ii) CNN, and (iii) RNN. 

Neural networks are non-linear models inspired by the neural architecture of the brain in biological systems. A typical neural network is known as a multi-layer perceptron and consists of a series of layers, composed of neurons and their connections \cite{Goodfellow2016}. 

\begin{figure}[ht!]
	\centering
	\includegraphics[width=0.95\linewidth]{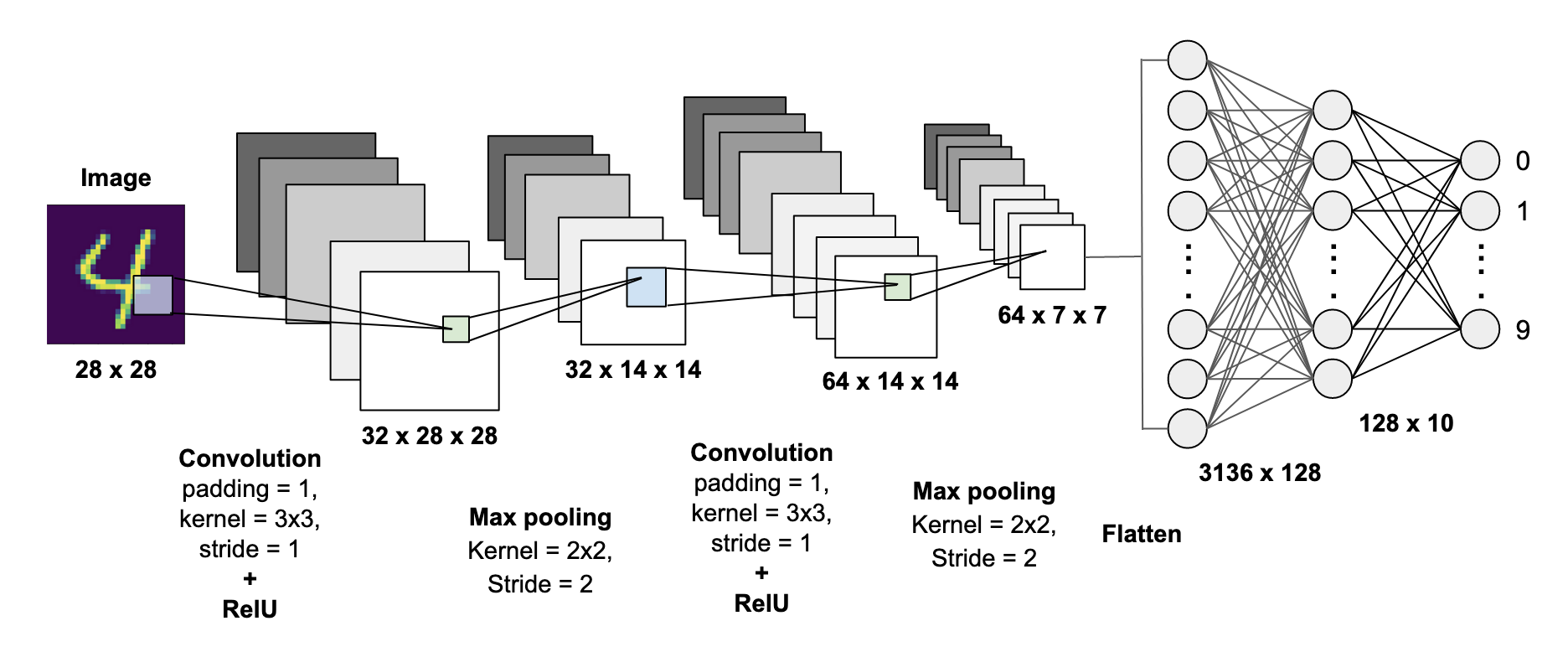}
	\caption[LeNet5 CNN architecture to classify handwritten characters.]{LeNet5 - LeCun et al.’s (1990) CNN architecture used to classify handwritten digits. This consists of two sets of convolutional and max pooling layers, followed by a flattening convolutional layer, then two fully-connected layers and finally a soft-max classifier.}
	\label{fig:CNN_arch}
\end{figure}

CNN are regularized version of MLPs with convolution operations in place of general matrix multiplication in at least one of the layers \cite{LeCun1989}. These types of networks find their motivation from work by \cite{Hubel1959, Hubel1962}. Inspired by this work, \cite{Fukushima1982} introduced convolutional layers and downsampling layers, \cite{Zhou1988} developed max pooling layers and \cite{LeCun1990} used back-propagation to derive the kernel coefficients for convolutional layers. The architecture of the NN used by LeCun et al. is known as LeNet5 and is shown in Figure~\ref{fig:CNN_arch} for the classification of hand-written digits. This essentially laid the foundations for modern CNNs. \cite{LeCun2010} gives a comprehensive history up to 2010 and a more recent review is available by \cite{Khan2020}.

RNN are in the family of feed-forward neural networks that have recurrent connections or allow for parameter sharing \cite{Lang1988}. These recurrent connection allow for input activations to pass from the hidden nodes in previous states to influence the current input \cite{Waibel1989, Lang1990}. 
% This is known as unrolling a RNN and is shown in Figure~\ref{fig:RNN_arch}.
% \clearpage
% \begin{figure}[ht!]
% 	\centering
% 	\includegraphics[width=0.95\linewidth]{unrolled_RNN.png}
% 	\caption[An unrolled RNN]{An unrolled RNN showing the recurrent connections between RNN cells $A$ for some input $x_t$ and outputs value $h_t$. From \cite{Olah2015}.}
% 	\label{fig:RNN_arch}
% \end{figure}
LSTM networks are one of the most commonly used variations of RNNs \cite{Patterson2017}. These were introduced by \cite{Hochreiter1997} and add the concept of memory gates or cells \cite{Graves2012, Gers1999}. These gates allow for information to be accessible across different time-steps and thus attenuates the vanishing gradient problem present with most RNN models \cite{Patterson2017}.

\subsection{Not Just Algorithms}
\begin{figure}[ht!]
	\centering
	\includegraphics[width=0.95\linewidth]{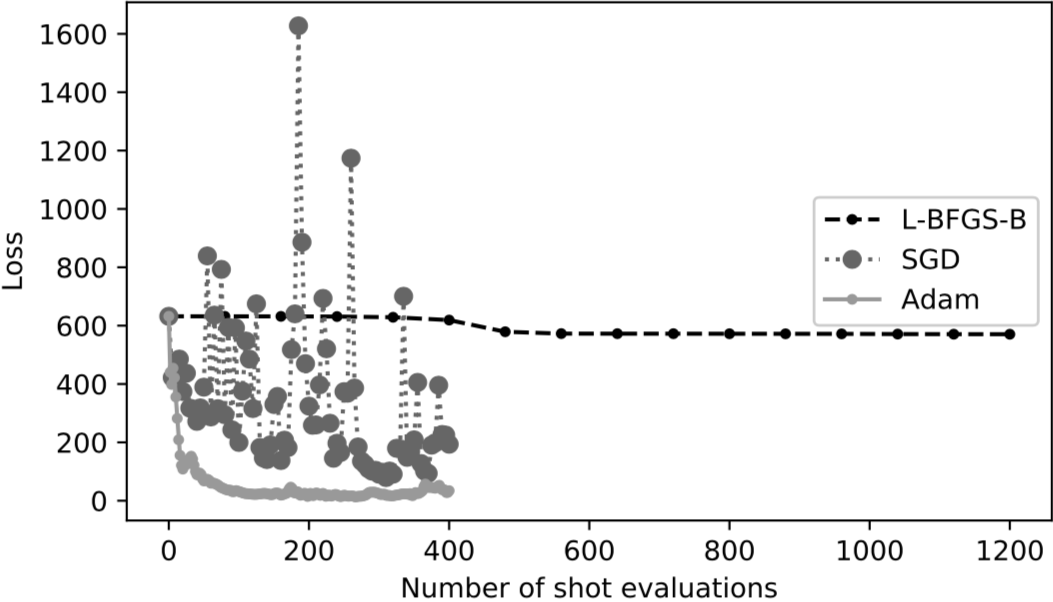}
	\caption[Faster convergence of Adam cost function.]{Adam cost function shown to converge much more rapidly than conventional Stochastic Gradient Descent and L-BFGS-B for FWI. From \cite{Richardson2018}.}
	\label{fig:state_of_art_adam}
\end{figure}

Apart from Machine Learning algorithms, re-interest in DNN has led to software architectures that allow for quick development. The most common include Tensorflow \cite{Abadi2015}, Keras \cite{Chollet2015}, PyTorch \cite{Paszke2017}, Caffe \cite{Jia2014}) and Deeplearning4j \cite{Nicholson2016}. These types of frameworks are facilitating interdisciplinarity between Machine Learning and geophysics. Indeed, \cite{Richardson2018} employed DNN architecture within Tensorflow to solve for FWI. Utilizing a common DNN optimizer - Adam - he shows in Figure~\ref{fig:state_of_art_adam} how the cost function converged quicker in the inversion process as compared to conventional methods in FWI. 

\section{Deep Learning in Geophysics}\label{sec:app_geophysics}
Machine Learning techniques have been utilised across different geophysical applications. Some notable examples include geo-dynamics \cite{Shahnas2018}, geology \cite{Reading2015}, seismology \cite{Shimshoni1998}, paleo-climatology \cite{Dowla1996}, climate change \cite{Anderson2018} and hydrogeology \cite{Hamshaw2018}. Unsupervised algorithms have also been investigated by \cite{Kohler2010} for pattern recognitions of wavefield patterns with minimal domain knowledge. Other geophysical application include seismic deconvolution \cite{Wang1992, CaldeOn-Macias1997}, tomography \cite{Nath1999}, first-break picking \cite{Murat1992}, trace editing \cite{McCormack1993a}, electricity \cite{Poulton1992}, magnetism \cite{Zhang1997}, shear-wave splitting \cite{Dai1994}, event classification \cite{Romeo1994}, petrophysics \cite{Downton2018} and noise attenuation \cite{Li2018a, Halpert2018}. 

\subsection{Legacy Velocity Inversion}\label{sec:dl_geo_leg_inv}
\begin{figure}[ht!]
	\begin{minipage}[c]{0.55\textwidth}
		\centering
		\includegraphics[width=\textwidth]{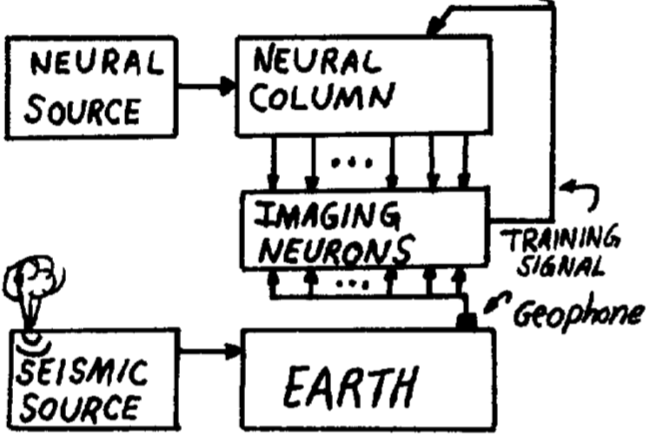}
	\end{minipage}\hfill
	\begin{minipage}[c]{0.4\textwidth}
		\caption[Block diagram for RNN system by Michaels \& Smith (1992)]{Block diagram for RNN system by \cite{Michaels1992}. The difference between a signal and the internal neural signals along a neural column are processed to provide a training signal that modifies neuron weights.}
		\label{fig:Block_RNN_FWI}
	\end{minipage}
\end{figure}

More specific to velocity estimation, the first published investigation for the use of NN was a RNN by \cite{Michaels1992}. Their network architecture represented all components in an elastic FWI experiment with a seismic source, the propagation media and an imaging response. Figure~\ref{fig:Block_RNN_FWI} shows a block diagram representation for their network. The neural column consisted of two 1-layer neuron columns, one for particle displacement and another for particle velocity. 

% \begin{figure}[ht!]
% 	\centering
% 	\includegraphics[width=0.6\linewidth]{RNN_FWI.png}
% 	\caption[Block diagram for RNN system by Michaels \& Smith (1992)]{Block diagram for RNN system by \cite{Michaels1992}. The difference between a signal and the internal neural signals along a neural column are processed to provide a training signal that modifies neuron weights.}
% 	\label{fig:Block_RNN_FWI}
% \end{figure}

\begin{figure}[ht!]
	\centering
	\includegraphics[width=0.95\linewidth]{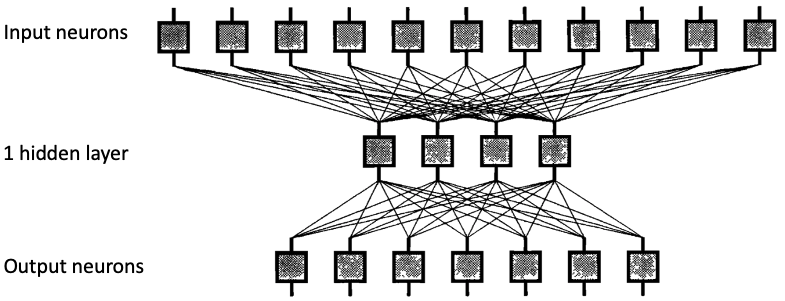}
	\caption[Architecture for first NN application to FWI.]{Architecture for first NN application to FWI. This was a very shallow NN with 1 hidden layer and non-symmetric input-output neurons. Adapted from \cite{Roth1994}.}
	\label{fig:first_NN_FWI}
\end{figure}

\cite{Roth1994} published the first application of NN which estimated 1D velocity functions from shot gathers from a single layer NN in 1994. Figure~\ref{fig:first_NN_FWI} shows their NN architecture. This accepted synthetic common shot gathers from a single source as input and used to compute corresponding 1D large-scale velocity models. The training set used for learning consisted of 450 synthetic models built up of eight strata with constant layer thickness over a homogeneous half-space. Their network was able to approximate a true velocity model sufficiently to act as a starting model for further seismic imaging algorithms. The inferred velocity profiles of the unseen data provided 80\% accuracy levels, and although the network was stable for noise contained data, it was not robust against strong correlated noise. Nonetheless, this investigation sets up NN as possible candidates to solve non-trivial inverse problems.

NNs are not solely limited to creating initial models to FWI. \cite{Langer1996} show how NNs can invert for parameters related to the seismic source and propagation medium using a similar single hidden layer architecture as R{\"o}th \& Tarantola. The difference in approach was two-fold; the NN employed a single seismogram as input as opposed to whole shot gather and pseudo-spectral data was used for training rather than time waveforms directly. The use of a single waveform did allow for improved results, however the use of pseudo-spectral data was instrumental. Transformed NN inference had better accuracies than the conventional time approach. Motivation to use pseudo-spectral data followed the work of \cite{Falsaperla1996} where they identified how the introduced sparsity within pseudo-spectral domain facilitated the learning process for the NN and was more robust to noise.

\subsection{Data-Driven Approaches to Velocity Estimation}\label{sec:dl_geo_data}
The terminology of data-driven geophysics is not a novel-one. This was first introduced in the literature by \cite{Schultz1994} when estimating rock properties directly from seismic-data through statistical techniques. However, conceptually, this is similar to the deconvolution process within a seismic processing flow \cite{Robinson1957,Robinson1967}. Namely, a filter is derived via autocorrelations and applied as a deconvolution operator \cite{Webster1978}. The term has only recently found a re-invigorated interest. Some modern applications of data-driven geophysical processes include dictionary learning \cite{NazariSiahsar2017}, time series analysis \cite{Wu2018a}, fault identification \cite{Mangalathu2020}, and reservoir characterization \cite{Schakel2014}.

Twenty-one years after \cite{Michaels1992}, \cite{Lewis2017} employed DNN architecture to learn prior models for seismic FWI. Their data driven approach at estimating initial models was applied to salt body reconstruction by learning the probability of salt geo-bodies and use this to regularize the FWI objective function. \cite{Araya-Polo2018} utilised DNN architecture and inverted for 2D high-velocity profiles. For the training process, they generated thousands of random 2D velocity models with up to four faults in them, at various dip angles and positions. Each model had three to eight layers, with velocities ranging from 2000 to 4000 \si{ms^{-1}}, with layer velocity increasing with depth. The DNN architecture is not defined in their paper, however when applied to unseen data with and without salt anomalies, their results achieved accuracies well above 80\% for both cases. This was used to obtain a low-wavenumber starting model then passed to traditional FWI as an initial model. \cite{Wu2018} proposed a convolutional-based network called ``InversionNet'' to directly map the raw seismic data to the corresponding seismic velocity model for a simple fault model with flat or curved subsurface layers. More recently, \cite{Li2019} extended this further and developed a DNN framework called ``SeisInvNet'' to perform the end-to-end velocity inversion mapping with enhanced single-trace seismic data as the input in time domain. 

\subsection{Wave Physics as an Analogue Recurrent Neural Network}\label{sec:dl_geo_theory}
\begin{figure}[ht!]
	\centering
	\includegraphics[width=0.98\linewidth]{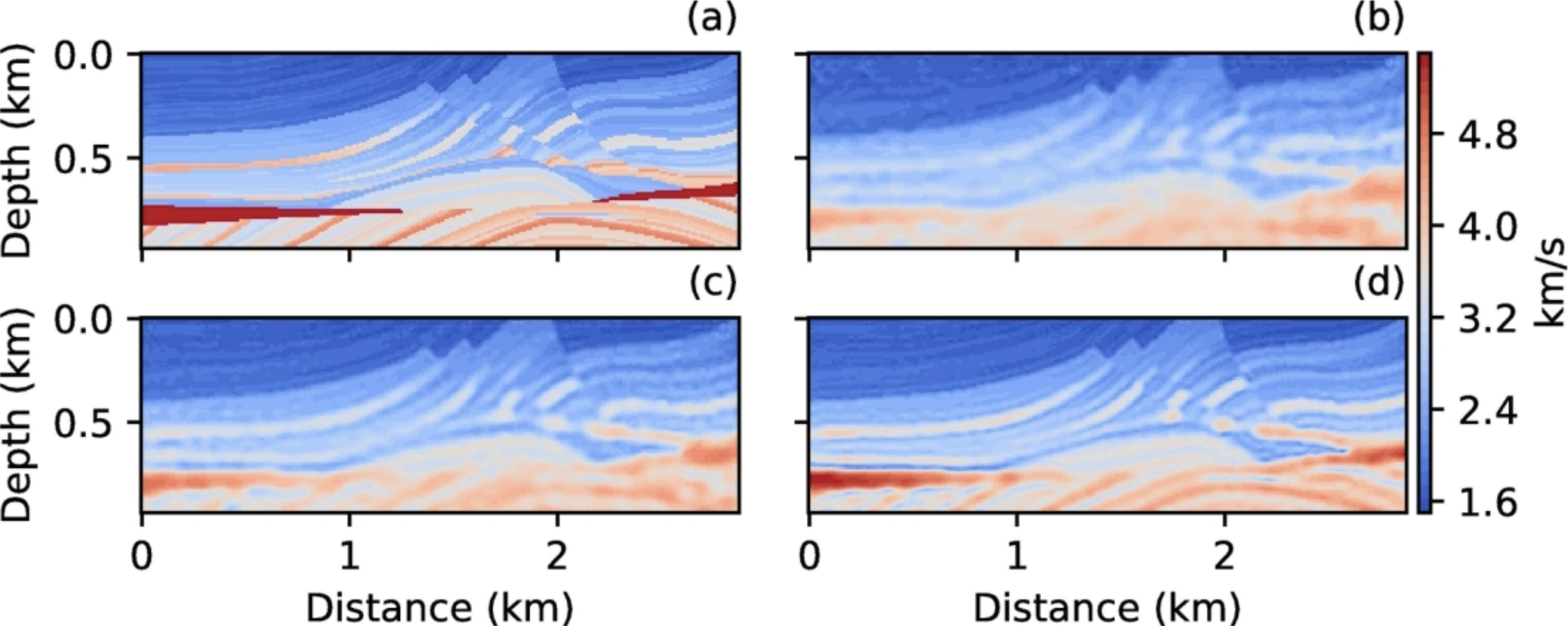}
	\caption[The inversion of Marmousi velocity model using RNN forward modelling.]{The inversion of Marmousi velocity model using RNN forward modelling framework with Adam algorithm optimizer. (a) True Marmousi. (b) 25th iteration. (c) 50th iteration (d) 100th iteration. From \cite{Sun2019}.}
	\label{fig:RNN_FWI_Marmousi}
\end{figure}

Recently, \cite{Raissi2019} and \cite{Hughes2019} derived a function between the physical dynamics of wave phenomena and RNNs. In their work they propose physics-informed neural networks that are trained to solve supervised learning tasks while respecting the laws of physics described by general non-linear partial differential equations. Fundamental to their approach is the ability for DNNs to be universal function approximators. Within this formulation, \cite{Raissi2019} are able to solve non-linear problems without the need to compute a priori assumptions, perform linearisation or employ local time-stepping. Under this new paradigm in modelling, \cite{Raissi2019} show how back-propagation is used ``to differentiate neural networks with respect to their input coordinates and model parameters to obtain physics-informed neural networks. Such NN are constrained to respect any symmetries, invariances, or conservation principles originating from the physical laws that govern the observed data, as modelled by general time-dependent and non-linear partial differential equations''. In particular, following up from this work, \cite{Sun2019} recast the forward modelling problem in FWI into a deep learning network by recasting the acoustic wave propagation into a RNN framework. Figure~\ref{fig:RNN_FWI_Marmousi} shows velocity inversion results from \cite{Sun2019} applied to the Marmousi velocity model. These theory-guided inversions still suffer from cycle-skipping, local-minima and high computational cost \cite{Sun2019}. Recent research suggests that Stochastic Gradient Descent algorithms have the capacity to escape local minima to a certain extent \cite{Sun2019}. 

% \begin{figure}[ht!]
% 	\centering
% 	\includegraphics[width=0.9\linewidth]{RAISSI_Wave_physics_analog_RNN.png}
% 	\caption[Solution to Schr{\'o}dinger's equation using RNN.]{Top: Predicted solution for the Schr{\'o}dinger equation $|h(t,x)|$ along with the initial and boundary training data.\newline Bottom: Comparison of the predicted and exact solutions corresponding to the three temporal snapshots depicted by the dashed vertical lines in the top panel. The relative L2 error is  $1.97\times10^{-3}$. From \cite{Raissi2019}.}
% 	\label{fig:RNN_physics}
% \end{figure}

\section{Challenges, Limitations, and Future Research Directions}\label{sec:challanges}

\subsection{Critical Challenges and Potential Limitations}

\begin{itemize}
    \item \textbf{Complexity and Model Size}: Deep learning models can be computationally intensive, requiring substantial computational resources and memory. The complexity of the models might hinder their practical application to large-scale FWI problems.
    
    \item \textbf{Data Requirements and Quality}: Deep learning models often demand large amounts of high-quality labeled data for training. In the context of FWI, obtaining such datasets might be challenging due to limited data availability and potential noise in seismic measurements.
    
    \item \textbf{Generalization and Overfitting}: Ensuring that deep learning models generalize well to unseen data is crucial. Overfitting to limited training data can lead to poor performance when faced with new geological settings or variations.
    
    \item \textbf{Interpretability and Uncertainty}: Deep learning models are often viewed as black boxes, which makes it challenging to interpret their decisions. Moreover, quantifying uncertainties in predictions and incorporating these uncertainties into FWI results remains a significant challenge.
    
    \item \textbf{Transferability}: Deep learning models trained on one geological setting might not easily transfer to another, due to variations in geological characteristics and seismic acquisition setups.
\end{itemize}

\subsection{Future Research Directions}

\begin{itemize}
    \item \textbf{Hybrid Models}: Investigate hybrid models that combine data-driven and theory-guided approaches, aiming to capitalize on the strengths of both paradigms. Developing techniques that effectively fuse data-driven learning with physical constraints could lead to more robust and accurate inversions.
    
    \item \textbf{Generative Models}: Explore the potential of generative models, such as Generative Adversarial Networks (GANs), for generating realistic synthetic seismic data. These models could supplement limited real data for training, potentially alleviating the data scarcity issue.
    
    \item \textbf{Uncertainty Quantification}: Research methodologies for estimating and quantifying uncertainties in the deep learning-based FWI results. Bayesian approaches, ensemble methods, and Monte Carlo techniques could be employed to provide a more comprehensive understanding of the inversion uncertainties.
    
    \item \textbf{Transfer Learning and Domain Adaptation}: Investigate techniques that enable the transfer of knowledge from one geological setting to another, adapting deep learning models to new environments with limited data.
    
    \item \textbf{Physics-Informed Learning}: Develop approaches that integrate domain-specific physical laws and constraints directly into deep learning architectures. Combining physics-based models with data-driven learning could enhance the accuracy and stability of inversions.
    
    \item \textbf{Sensitivity Analysis}: Explore methods to analyze the sensitivity of deep learning models to different input data types, noise levels, and model hyperparameters. Understanding how variations impact the model's performance can lead to more robust designs.
    
    \item \textbf{Incorporating Multiple Modalities}: Investigate methods to incorporate diverse types of data sources, such as well logs, gravity data, or electromagnetic data, into the FWI process. Integrating multi-modal data could improve inversion accuracy and provide more comprehensive subsurface characterizations.
    
    \item \textbf{Collaborative Frameworks}: Foster collaboration between deep learning researchers, geophysicists, and domain experts to jointly develop and validate new approaches. Cross-disciplinary efforts can lead to innovative solutions that address the unique challenges of FWI.
\end{itemize}

In conclusion, while the integration of deep learning and FWI holds immense promise, there are several critical challenges and potential limitations that must be addressed. The outlined future research directions aim to guide the evolution of this field towards more robust, accurate, and practically applicable solutions. By addressing these challenges and exploring innovative directions, researchers can unlock the full potential of deep learning in revolutionizing the field of seismic imaging and subsurface characterization.

\section{Summary}\label{sec:summary}

This review paper explores the integration of deep learning techniques into the field of full-waveform inversion (FWI) for seismic imaging and subsurface characterization. The paper comprises five chapters that cover various aspects of this integration, including the theoretical foundations, application examples, challenges, limitations, and future research directions. The main findings and discussions from each chapter are summarized as follows:

\subsection{Chapter 1: Introduction}
We introduced the motivation behind integrating deep learning with FWI and provided an overview of the paper's structure. We highlighted the potential of deep learning to address the challenges associated with traditional FWI methods and outlined the objectives of the paper.

\subsection{Chapter 2: Fundamentals of Full-Waveform Inversion}
This chapter delved into the theoretical foundations of FWI. We covered the mathematical principles underlying FWI, discussed the forward and inverse problems, and explained the key components of FWI workflows. The importance of accurately estimating subsurface properties through seismic data inversion was emphasized.

\subsection{Chapter 3: Deep Learning Fundamentals}
We provided a comprehensive introduction to deep learning techniques. The chapter covered essential concepts such as neural networks, activation functions, loss functions, optimization algorithms, and regularization techniques. Understanding these fundamentals is crucial for grasping the subsequent chapters' applications in the context of FWI.

\subsection{Chapter 4: Deep Learning Applications in Geophysics}
This chapter explored specific applications of machine learning and deep learning techniques in geophysics. Notable examples included various geophysical domains like seismology, geology, and hydrogeology. We detailed how deep learning methods have been applied to address problems such as velocity estimation, seismic deconvolution, and tomography, showcasing the potential for enhanced subsurface characterization.

\subsection{Chapter 5: Challenges, Limitations, and Future Research Directions}
We highlighted critical challenges and potential limitations associated with integrating deep learning and FWI. These challenges include model complexity, data quality and quantity, generalization, interpretability, and transferability. The chapter then outlined various future research directions to address these challenges. Hybrid models, generative models, uncertainty quantification, transfer learning, and physics-informed learning were among the proposed strategies for advancing the field.

In conclusion, this review paper underscores the significant potential of integrating deep learning techniques into the realm of FWI for seismic imaging and subsurface characterization. Through an exploration of theoretical foundations, practical applications, challenges, and future research directions, it becomes evident that while this integration presents challenges, it also offers promising avenues for enhanced accuracy, efficiency, and reliability in subsurface property estimation. By addressing these challenges and pursuing innovative research directions, the synergy of deep learning and FWI holds the promise of revolutionizing the field of geophysics and contributing to a deeper understanding of Earth's subsurface properties.

\bibliographystyle{unsrt}  
\bibliography{references}  %%% Remove comment to use the external .bib file (using bibtex).

\end{document}